\documentclass[twocolumn,prl,superscriptaddress,aps,showpacs]{revtex4}

\usepackage{amsmath}
\usepackage{amssymb}
\usepackage{amsfonts}
\usepackage{bm}
\usepackage{bbm}
\usepackage{epsf}
\usepackage{epsfig}
\usepackage{times}
\usepackage{nicefrac}
\usepackage{graphicx}
\usepackage{maybemath}
\def\half{{\textstyle{\frac12}}}

\def\dd{{\mathrm{d}}}
\def\ii{{\mathrm{i}}}

\def\SS{{\mathcal{S}}}

\def\calC{{\mathcal{C}}}

\def\half{   {\textstyle{\frac12}}   }

\def\tfrac#1#2{ {\textstyle{\frac{#1}{#2}} } }
\def\trunc#1{[\mkern - 2.5 mu  [#1] \mkern - 2.5 mu ]}

\newcommand{\PT}{${\mathcal{PT}}$}

\begin{document}

\newcommand{\borel}[1]{\mbox{$\mathcal{#1}$}}

\newcommand{\addrHD}{Max--Planck--Institut f\"ur Kernphysik,
Postfach 103980, 69029 Heidelberg, Germany}

\newcommand{\addrRolla}{
Department of Physics, Missouri University of Science
and Technology, Rolla MO65409-0640, USA}

\newcommand{\addrHDphiltheo}{Institut f\"ur Theoretische Physik,
Universit\"{a}t Heidelberg,
Philosophenweg 16, 69120 Heidelberg, Germany}

\newcommand{\addrHDphilphys}{Physikalisches Institut der Universit\"{a}t,
Philosophenweg 12, 69120 Heidelberg, Germany}

\newcommand{\addrSACLAY}{CEA, IRFU and Institut de Physique
Th\'{e}orique, Centre de Saclay, F-91191 Gif-Sur-Yvette, France}

\title{Unified Treatment of Even and Odd Anharmonic Oscillators of
Arbitrary Degree}

\author{Ulrich D.~Jentschura}\email{ulj@mst.edu}
\affiliation{\addrRolla}
\affiliation{\addrHD}
\author{Andrey Surzhykov}
\affiliation{\addrHD}
\affiliation{\addrHDphilphys}
\author{Jean Zinn-Justin}
\affiliation{\addrSACLAY}

\date{\today}

\begin{abstract}
We present a unified treatment, including higher-order
corrections, of anharmonic oscillators of arbitrary 
even and odd degree. Our approach is based on a dispersion
relation which takes advantage of the \PT{}-symmetry 
of odd potentials for imaginary coupling parameter,
and of generalized quantization conditions which take into 
account instanton contributions. We find a number of 
explicit new results, including the general behaviour 
of large-order perturbation theory for arbitrary levels
of odd anharmonic oscillators, and subleading corrections
to the decay width of excited states for odd potentials,
which are numerically significant.
\end{abstract}


\maketitle

{\em Introduction.---}One-dimensional anharmonic 
oscillators are quite basic.
Because of their enormous phenomenological significance,
they occupy a unique position within quantum 
theory~\cite{BeWu1969,BeWu1971,BeWu1973,BrLGZJ1977prd1,BrLGZJ1977prd2}.
They are treated at various levels 
of sophistication in nearly every textbook on quantum mechanics.
Here, we use the Hamiltonians of even oscillators in the convention
\begin{equation}
\label{HN}
H_N(g) = -\half \, \partial_q^2 + \half \, q^2 + g\, q^N \,, \qquad
\mbox{($N$ even)}\,,
\end{equation}
and odd oscillators as follows,
\begin{equation}
\label{hM}
h_M(g) = -\half \, \partial_q^2 + \half \, q^2 + \sqrt{g}\, q^M \,. \qquad 
\mbox{($M$ odd)}\,.
\end{equation}
For $g <0$, the potential of an even oscillator
has a double-hump structure, and it
is intuitively clear that the particle can tunnel 
through the barrier(s). 
This is manifest in the energy levels, because they
develop a nonvanishing imaginary part as we vary 
the coupling from positive $g$ to negative $g$ in the 
complex plane.
The smaller we choose the modulus of $g$, the bigger the humps,
the longer the tunneling time of the particular,
and the smaller is the decay width of the 
state (i.e., the smaller is the modulus of the imaginary part of the 
resonance energy). Indeed, the imaginary part has 
a ``nonperturbative'' behaviour in the coupling and is 
well known to be proportional to $\exp(-A/|g|^b)$, 
where $A$ and $b$ are positive constants.
The same is true for the odd oscillators for $g > 0$;
the resonance energies are manifestly complex and their
imaginary parts also involve non-analytic exponentials.

Given the large amount of work already invested by the 
physics community into the study of anharmonic oscillators,
it is perhaps surprising that 
two very basic basic questions regarding the above mentioned 
anharmonic oscillators have not yet been fully addressed
in the literature:
(i) What are the higher-order corrections to the 
nonperturbative behaviour of the resonance energies,
and how are the real and the imaginary part of the
resonance energy described by a (possibly) non-analytic,
generalized expansion in $g$? Which mathematical
structures (exponentials, logarithms, $\dots$) form part
of such an expansion?
(ii) What is the general large-order behaviour of 
perturbation theory for an arbitrary energy levels of an 
odd oscillator of arbitrary degree? 
Both of these questions are addressed here, and 
we also report on numerical calculations to test our 
analytic formulas, with special emphasis on the
anharmonic oscillators of degrees $6$ and $7$.

Indeed, to answer the above questions, we rely in part on 
the work of Bender and Wu who, in 1971 (see Ref.~\cite{BeWu1971}), 
solved question (ii) for even anharmonic oscillators, and on the 
concept of 
\PT{}-symmetry~\cite{BeBo1998,BeDu1999,BeBoMe1999,BeBrJo2002} 
for the formulation of 
a dispersion relation for the resonance energies of odd oscillators.
Another essential ingredient of our analysis are generalized quantization
conditions which allow us to describe instanton
contributions and which go beyond the ordinary Bohr--Sommerfeld 
formalism.

\begin{figure}[t]
\begin{center}
\begin{center}
\begin{minipage}[b]{1.0\linewidth}
\includegraphics[width=0.77\linewidth,angle=0, clip=]{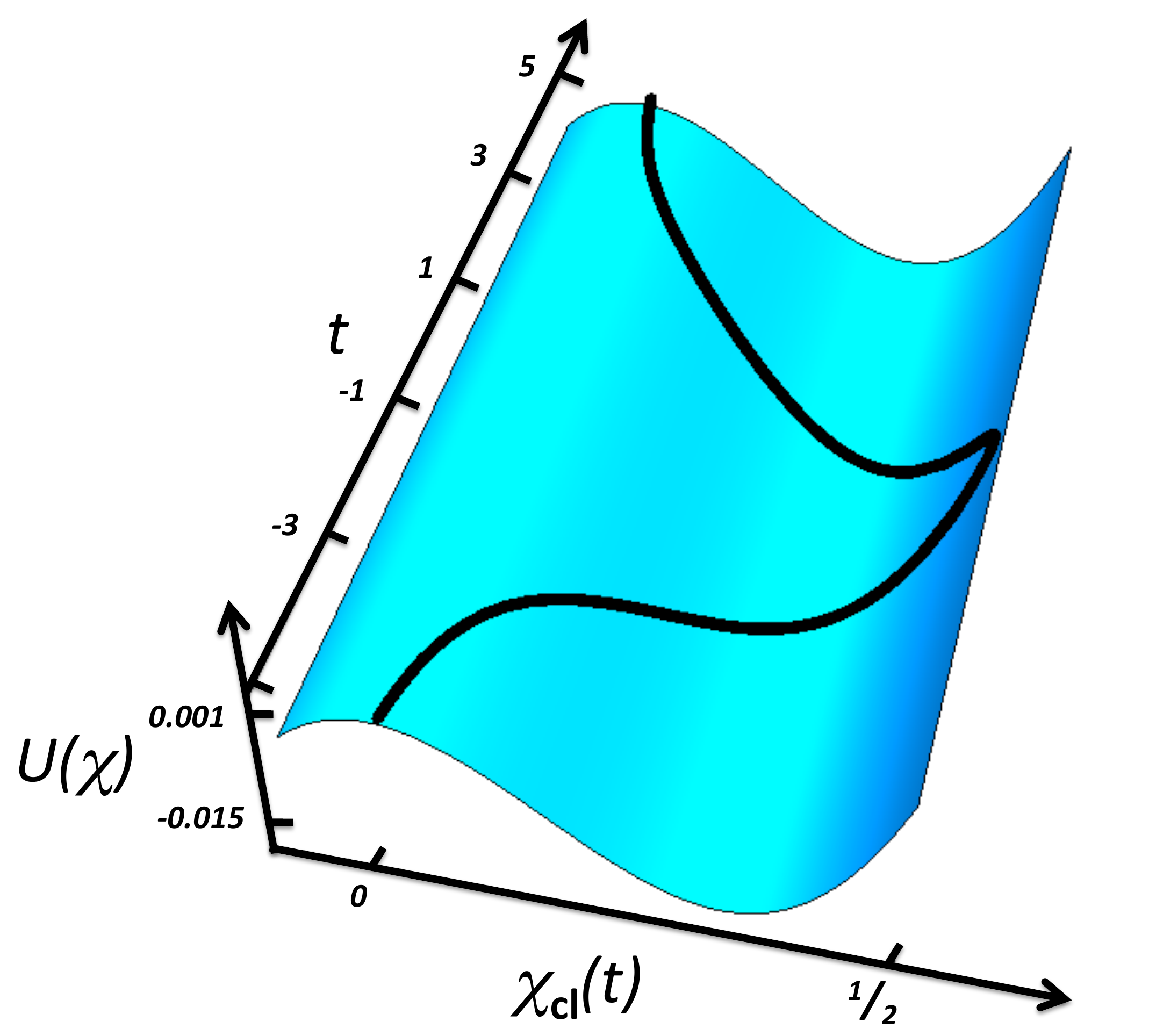} 
\end{minipage}
\begin{minipage}[b]{1.0\linewidth}
\caption{\label{Fig1} (color online.)
Instanton world-line configuration
for the cubic potential.
The plot shows the instanton world-line
$\chi_{\rm cl}(t) = [\cosh(t) + 1]^{-1}$
immersed in the scaled
potential $U(\chi) = \chi^3 - \half \, \chi^2$.}
\end{minipage}
\end{center}
\end{center}
\end{figure}

{\em Instanton actions.---}We 
consider even anharmonic oscillators
of degree $N$ in the convention~(\ref{HN})
with energy eigenvalues $E_n^{(N)}(g)$, and odd Hamiltonians in the 
convention (\ref{hM}) 
with complex resonance energies are $\epsilon_n^{(M)}(g)$.
Formulating the problem of the determination of energy 
levels in terms of a Euclidean path integral~\cite{ZJ1996},
it becomes clear that instanton configurations 
should be analyzed,
and we start with the case of even
oscillators. Here, the instanton
configuration exists for negative $g$, 
and we thus scale 
$q(t) = (-g)^{-1/(N-2)} \, \xi(t)$. The Euclidean action reads
\begin{equation}
\label{Seven}
S[\xi] = (-g)^{-\frac{2}{N-2}} \, 
\int \dd t \, \left( \half \, {\dot \xi}^2 + \half \, \xi^2 -
\xi^N \right) \,.
\end{equation}
The instanton configurations are ($N$ even)
\begin{equation}
\label{instantonEven}
q^{\pm}_{\rm cl}(t) = \pm 
(-g)^{-\frac{2}{N-2}} \, 
\left\{ 1 + \cosh[(N-2) (t - t_0)] \right\}^{-\frac{1}{N-2}} \,.
\end{equation}
Here, $t_0$ is a collective coordinate.
For odd anharmonic oscillators, we transform
$q(t) = - g^{-1/(2 M-4)} \chi(t)$ and obtain the Euclidean
action ($M$ odd)
\begin{equation}
\label{Sodd}
S'[\chi] = 
g^{-\frac{1}{M-2}} \, 
\int \dd t \, \left( \half \, {\dot \chi}^2 + \half \, \chi^2 -
\chi^M \right) \,.
\end{equation}
The instanton $q(t) = q_{\rm cl}(t)$ is unique 
because the potential has lost the invariance under parity
(see Fig.~\ref{Fig1}), 
\begin{equation}
\label{instantonOdd}
q_{\rm cl}(t) = 
g^{-\frac{1}{M-2}} \, 
\left\{ 1 + \cosh[(M-2) (t-t_0)] \right\}^{-\frac{1}{M-2}} \,.
\end{equation}
Inserting the solutions 
$q^\pm_{\rm cl}(t)$ and $q_{\rm cl}(t)$ into Eqs.~(\ref{Seven}) and 
(\ref{Sodd}), we obtain the classical Euclidean instanton actions
\begin{subequations}
\label{actionsEvenOdd}
\begin{align}
S[q^\pm_{\rm cl}] =& \left(- g\right)^{-\frac{2}{N-2}} 
{\cal A }(N), \;\;\;
S'[q_{\rm cl}] = g^{-\frac{1}{M-2}} 
{\cal A }(M),
\\[0ex]
\label{definitionA}
{\cal A }(m) =& \; 2^{2/(m-2)} \, 
B\left(  \frac{m}{m-2}, \frac{m}{m-2} \right) \,,
\end{align}
\end{subequations}
where $B(x, y) = \Gamma(x)\,\Gamma(y)/\Gamma(x + y)$ is the 
Euler Beta function.

%
%

{\em Dispersion relations.---}An evaluation of 
the quantum fluctuations about the instanton
configurations according to Ref.~\cite{ZJ1996} 
reveals that the imaginary part of the $n$th resonance energy
$E_n^{(N)}(g)$, for potentials of even order $N$,
is given by (in leading order, 
corrections are of relative order $g^{2/(N-2)}$)
\begin{align}
\label{ImEn}
& {\rm Im} \, E_n(N,g < 0) =  
-\frac{1}{n! \sqrt{2 \pi}} \, 
\left( -\frac{2 \calC(N)}{(-g)^{2/(N-2)}} \right)^{n + 1/2} 
\nonumber\\[0ex]
& \qquad \times
\exp\left( -(-g)^{-2/(N-2)} \, {\cal A}(N) \right) \,,
\end{align}
with $\calC(m) = 2^{2/(m-2)}$.
For odd potentials in the normalization given by
Eq.~(\ref{hM}) and for positive coupling,
ignoring corrections of relative order $g^{1/(M-2)}$,
\begin{align}
\label{Imepsn}
& {\rm Im} \, \epsilon_n^{(M)}(g > 0) =
-\frac{1}{2 n! \sqrt{2 \pi}} \, 
\left( \frac{2 \calC(M)}{g^{1/(M-2)}} \right)^{n + 1/2} 
\nonumber\\[0ex]
& \qquad \times
\exp\left( -g^{-1/(M-2)} \, {\cal A}(M) \right) \,.
\end{align}

The subtracted dispersion 
relation~\cite{LoMaSiWi1969,BeWu1971,BeWu1973} for the energies $E_n(N,g)$ 
of the even anharmonic oscillators of degree $N$ is 
\begin{equation}
\label{dispersionEven}
E_n^{(N)}(g) = n + \frac12 - \frac{g}{\pi}\, \int_{-\infty}^0 \dd s \,
\frac{{\rm Im}\, E_n^{(N)}(s+{\rm i}\,0)}{s \, (s - g)} \,.
\end{equation}
Using the formula (\ref{ImEn}) and the dispersion 
relation (\ref{dispersionEven}), one may calculate~\cite{BeWu1971}
the large-order perturbative expansion
for the ground-state energy of the $n$th level of an anharmonic
oscillator of order $N$, where we use the perturbative coefficients
in the form 
$E_n^{(N)}(g) \sim  \sum_K E_{n,K}^{(N)} \, g^K$.
For an arbitrary level of an even oscillator of arbitrary degree, 
we thus rederive~\cite{BeWu1971}
\begin{align}
\label{leadingEven}
E_{n,K}^{(N)} \sim & \; \frac{ (-1)^{K+1}\, (N-2)}%
{\pi^{3/2} \, n! \, 2^{K + 1 - n}} 
\Gamma\left( \frac{N - 2}{2} K + n + \half \right) 
\nonumber\\[0ex]
& \times \left[ B\left(\frac{N}{N-2}, \frac{N}{N-2}\right) 
\right]^{-\frac{N-2}{2} K - n - \half} \,,
\end{align}
which is valid up to corrections of relative order $K^{-1}$.
For odd Hamiltonians $h_M(g)$ which involve a perturbation 
of the form $\sqrt{g}\, q^M$, we have only one
branch cut in the energy~\cite{BeDu1999}.
The $\mathcal{PT}$-symmetry for purely imaginary coupling
leads to the following dispersion relation~\cite{BeDu1999,BeWe2001},
\begin{equation}
\label{dispersionOdd}
\epsilon_n^{(M)}(g) = n + \frac12 + 
\frac{g}{\pi}\, \int_0^{\infty} \dd s \,
\frac{{\rm Im} \, \epsilon_n^{(M)}(s + {\rm i}\,0)}{s\, (s - g)}\,.
\end{equation}
Based on the dispersion relation (\ref{dispersionOdd}) and on the general 
result for the imaginary part of the resonance energy 
for an odd anharmonic oscillator given in Eq.~(\ref{Imepsn}),
we are now in the position to write down 
the large-order behaviour of the 
perturbative coefficients of an arbitrary level of 
an odd anharmonic oscillator of arbitrary degree.
Specifically, for a resonance 
$\epsilon^{(M)}_n(g) \sim \sum_K \epsilon^{(M)}_{n,K} \, g^K$, 
we find ($M$ odd, $M \geq 3$)
\begin{align}
\label{leadingOdd}
\epsilon^{(M)}_{n,K} \sim & \; \frac{2 - M}{\pi^{3/2} \, n! \, 2^{2 K + 1 - n}} 
\Gamma\left( (M - 2) K + n + \half \right)
\nonumber\\[0ex]
& \times \left[ B\left(\frac{M}{M-2}, \frac{M}{M-2}\right) 
\right]^{-(M - 2) K - n - \half}  \,.
\end{align}

{\em Subleading
corrections.---}In order to go beyond the leading-order
results, further considerations are needed.
While details of the derivation
will be presented in~\cite{JeSuZJ2009prep},
we would like to give here the essential ingredients
of our formalism. First of all, we scale the 
coordinate in the 
Hamiltonians (\ref{HN}) and (\ref{hM}) as 
$q \to g^{-\frac{2 \trunc{m/2} - m + 2}{m-2}} \, q$,
where $\trunc{X}$ is the integer part of $X$
($m$ stands for a general integer).
We then transform the Schr\"{o}dinger to the Riccati
equation by setting
$\varphi'(q)/\varphi(q) = - 
S(q)\, g^{-\frac{2 \trunc{m/2} - m + 2}{m-2}}$,
and we denote by $S_+(q)$ the component of 
$S(q)$ which is even under the 
operation $(g, E) \to (-g, -E)$.
For $S_+$, we calculate the Wentzel--Kramers--Brioullin (WKB)
expansion for $S(q)$ for given $g$ and energy $E$,
using the algorithm described for a general potential
in Sec.~3 of Ref.~\cite{ZJJe2004i},
by expanding the solution in fractional powers of $g$ 
while keeping the quantity
$g^{\frac{2 \trunc{m/2} - m + 2}{m-2}} E$ fixed.
A recursive procedure for the construction of the WKB
expansion of $S_+(z, g, E)$ has been outlined in 
Eqs.~(3.40)---(3.42) of Ref.~\cite{ZJJe2004i}.
We then integrate the WKB expansion of the function 
$S_+(q)$ around its cuts 
using an approach based on Mellin transforms 
as outlined in Appendix F.3 of Ref.~\cite{ZJJe2004ii}
for general potential [the integration contour 
${\mathcal C}'$ 
around the cuts is
chosen in accordance with Eq.~(3.52) of Ref.~\cite{ZJJe2004i}]
and conjecture the following form for the result
of the contour integral,
\begin{align}
\label{genie}
& g^{-\frac{2 \trunc{m/2} - m + 2}{m-2}}
\oint_{\mathcal C'} \dd z \, \SS_+(q) = 
A_m(E, g) + \ln(2 \pi) 
\\[0ex]
& -\ln \Gamma\left( \half - B_m(E, g) \right)
+ B_m(E, g) \, \ln\left( - \frac{g}{2 \, {\mathcal C}(m)} \right) \,.
\nonumber
\end{align}
From the right-hand side, under a suitable separation of
real and imaginary parts, one can directly read off the 
two functions $B_m(E,g)$ and $A_m(E,g)$,
which we refer to as the ``perturbative function''
and the ``instanton function,'' respectively.
We here
indicate for completeness the first few terms of $B_m(E,g)$ for the 
oscillators with $M = 3,7$ and $N = 4,6$, 
\begin{subequations}
\label{Bdata}
\begin{align}
\label{B3}
B_3(E,g) =& \; E + g \left( \tfrac{7}{16} + 
\tfrac{15}{4} \, E^2 \right) 
+ {\mathcal O}(g^2) \,, \\
\label{B4}
B_4(E, g) =&\; E - g \left( \tfrac{3}{8} + 
\tfrac{3}{2} \, E^2 \right) 
+ {\mathcal O}(g^2) \,, \\
\label{B6}
B_6(E, g) =&\; E - g \left( \tfrac{25}{8} \, E + 
\tfrac{5}{2} \, E^3 \right) 
+ {\mathcal O}(g^2) \,,\\
B_7(E, g) =&\; E + g \left( \tfrac{180675}{2048} + 
\tfrac{444381}{512} \, E^2 \right. 
\nonumber\\
& \; \left. + \tfrac{82005}{128} \, E^4 
+ \tfrac{3003}{32} \, E^6 \right) 
+ {\mathcal O}(g^2) \,.
\end{align}
\end{subequations}
The leading term of the $A$ functions contains the instanton 
action as given in Eq.~(\ref{actionsEvenOdd}):
$A_N(E,g) = {\cal A }(N)\,(-g)^{-2/(N-2)} + 
{\mathcal O}(g^{2/(N-2)})$
for even potentials and
$A_M(E,g) = {\cal A }(M)\,g^{-1/(M-2)} + 
{\mathcal O}(g^{1/(M-2)})$
for odd potentials, respectively.
Higher-order terms read
\begin{subequations}
\label{Adata}
\begin{align}
\label{A3}
& A_3(E,g) = \tfrac{2}{15\,g} + 
g\, \left( \tfrac{77}{32} + \tfrac{141}{8} \, E^2 \right) 
+ {\mathcal O}(g^2) \,,
\end{align}\\[-8.0ex]
\begin{align}
\label{A4}
& A_4(E,g) = -\tfrac{1}{3\,g} - 
g\, \left( \tfrac{67}{48} + \tfrac{17}{4} \, E^2 \right) 
+ {\mathcal O}(g^2) \,,
\end{align}\\[-8.0ex]
\begin{align}
\label{A6}
& A_6(E,g) =
\tfrac{ \pi }{2^{5/2} \, (-g)^{1/2}} 
- g \, \left( \tfrac{221}{24} \, E + \tfrac{17}{3} \, E^3 \right) 
\\[0ex]
& + g^2 \, \left( \tfrac{ 2504899 }{ 7680 } \, E + 
\tfrac{ 45769 }{ 96 } \, E^3 +
\tfrac{ 17527 }{ 160 } \, E^5 \right) + {\mathcal O}( g^3 ) \,,
\nonumber
\end{align}\\[-8.0ex]
\begin{align}
\label{A7}
& A_7(E,g) =
\frac{5^{1/4} \, \Gamma(\tfrac{1}{5} )\,\Gamma(\tfrac{2}{5})}%
  {2^{3/5} \, 9 \, \pi \, \sqrt{\phi} \, g^{1/5}} 
\\[0ex]
& \quad + g^{1/5} \frac{5^{1/4} \, 
\Gamma^2( \tfrac{3}{5} )\,\Gamma( \tfrac{4}{5} )}%
{2^{7/5} \, \pi \, \sqrt{\phi} } \,
\left(\tfrac{5}{8} + \tfrac{9}{10} \, E^2 \right) + {\cal O}(g^{2/5})\,.
\nonumber
\end{align}
\end{subequations}
The function $A_7(E,g)$ involves 
the square root of the golden ratio $\phi = (\sqrt{5}+1)/2$.
The result for the sextic oscillator ($N=6$) is at variance with 
expectation because one would have assumed the presence 
of a term of order $g^{2/(N-2)} = g^{1/2}$, but it cancels, accidentally.

In terms of the $A$ and $B$ functions, 
we can now write down our conjectures for the 
generalized quantization conditions.
For the ``stable'' cases (odd oscillator,
imaginary coupling and even oscillator,
positive coupling), we simply conjecture them to read
(we indicate two equivalent forms)
\begin{equation}
\label{pertquant}
1/\Gamma[ \half - B_m(E,g) ] = 0\,,\quad
B_m(E,g) = n + \tfrac12\,,
\end{equation}
where $m$ can be even or odd.
This form of the condition
generalizes the result $1/\Gamma(\tfrac12 - E) = 
{\rm det}(H - E) = 0$ for the Fredholm determinant
of the harmonic oscillator Hamiltonian $H$.
In the presence of instanton configurations,
the resonance energy is 
slightly displaced from the energy that would otherwise
lead to a pole of the $\Gamma$ function according to
Eq.~\eqref{pertquant}. The displacement of $E$ is 
by a nonperturbative correction which can be 
evaluated exactly in leading order [see Eqs.~\eqref{ImEn}
and~\eqref{Imepsn}]; this means that the zero on the right-hand side
of $1/\Gamma[ \half - B_m(E,g) ] = 0$ has to be replaced
by the nonperturbatively small imaginary part of the resonance energy.
A comparison of the resulting equation to the
leading terms of the functions $A_m(E,g)$ and $B_m(E,g)$ which 
emerge from the evaluation of the contour integral
of the WKB expansion \eqref{genie}
then, in turn, naturally leads to the following 
conjectures for even potentials ($g < 0$) and 
odd potentials ($g > 0$), respectively, 
\begin{subequations}
\label{EvenOddQuant}
\begin{align}
\label{EvenQuant}
& \frac{\Gamma\left( \half - B_N(E,g) \right)}%
{\sqrt{2 \pi} \,\, {\rm e}^{A_N(E, g)}}
\left(-\frac{2 \, \calC(N)}{(-g)^{\frac{2}{N-2}}} \right)^{B_N(E, g)} 
=1 \,,\\[1ex]
& \label{OddQuant}
\frac{\Gamma\left( \half - B_M(E,g) \right)}%
{\sqrt{8 \pi} \,\, {\rm e}^{A_M(E, g)}} 
\left( \frac{2 \, \calC(M)}{g^{\frac{1}{M-2}}} \right)^{B_M(E, g)} \!\!\!
=1.
\end{align}
\end{subequations}
In order to solve the perturbative quantization 
condition~\eqref{pertquant}, we enter with an ansatz
$E = E_0 + g E_1 + g^2 E_2 + \dots$ and 
compare coefficients in each order in $g$.
In order to solve \eqref{EvenOddQuant}, our ansatz 
also has involve nonanalytic terms as implied by the 
instanton contributions, and we then expand systematically
in powers of $g$ and simultaneously in powers of the nonanalytic
factor $\exp(-A/|g|^p)$ (with $A$ and $p$ suitably chosen).
This ansatz naturally leads to the triple expansions
[with constant coefficients $\Xi^{(m,n)}_{J,L,K}$ and 
$L_\mathrm{max} \equiv {\mathrm{max}}(0,J-1)$],
\begin{widetext}
\begin{subequations}
\label{evenodd}
\begin{align}
\label{even}
E_n^{(N)}(g < 0) = &
\sum_{J=0}^\infty 
\left[ 
\left( \frac{2 {\cal C}(N)}{(-g)^{2/(N-2)}} \right)^{n + \half} 
\frac{\ii \exp\left(- \frac{{\cal A}(N)}{(-g)^{2/(N-2)}}\right) }%
{n! \sqrt{2 \pi}} \right]^J
\sum_{L=0}^{L_\mathrm{max}}
\ln^L\left( -\frac{ 2 {\mathcal C}(N) }{ (-g)^{2/(N-2)}}  \right)
\sum_{K=0}^\infty
\Xi^{(N,n)}_{J,L,K} 
(-g)^{2 K/(N-2)} ,
\end{align}
\begin{align}
\label{odd}
\epsilon_n^{(M)}(g > 0) = & 
\sum_{J=0}^\infty 
\left[ 
\left( \frac{2 {\cal C}(M)}{g^{1/(M-2)}} \right)^{n + \half} \,
\frac{\ii \exp\left(-\frac{{\cal A}(M)}{g^{1/(M-2)}}\right)}%
{n! \, \sqrt{8 \pi}}
\right]^J
\sum_{L=0}^{L_\mathrm{max}}
\, \ln^L\left( -\frac{ 2 {\mathcal C}(M) }{ g^{1/(M-2)}}  \right)
\sum_{K=0}^\infty
\Xi^{(M,n)}_{J,L,K} \, g^{K/(M-2)} \,.
\end{align}
\end{subequations}
\end{widetext}
The above triple expansion is involved and in need of an interpretation
[we concentrate on the ``even case'' (\ref{even})].
The term with $J = 0$ recovers the basic, perturbative expansion 
which has only integer powers in $g$. Therefore, 
$\Xi^{(N,n)}_{0,0,K(N-2)/2} = (-1)^K \, E^{(N)}_{n,K}$.
The term with $J = 1$ recovers the leading contribution
in the expansion in powers of $\exp[-{\mathcal A}(N)/(-g)^{2/(N-2)}]$ 
to the imaginary part of the resonance energy,
but including perturbative corrections which can be
expressed in terms of a fractional power series in $g$
multiplying the nonanalytic exponential.
The first few terms of this series
are given below for the first excited state of the cubic 
potential.
The term with $J = 2$ involves a logarithm of the 
form $\ln\left( -\frac{ 2 {\mathcal C}(N) }{ (-g)^{2/(N-2)}} \right)$.
The explicit imaginary parts of the logarithms cancel 
against the imaginary parts of
Borel sums carried out in complex directions~\cite{complex} 
of perturbation series that occur in lower-order (in $J$) contributions.
Indeed, the Laplace--Borel integrals have to be carried out 
in a manner consistent with the analytic continuation of the 
logarithms~\cite{JeZJ2001,ZJJe2004ii}.

The above formulas allow, in principle, the determination
of corrections of arbitrary order to the resonance energies
of oscillators of arbitrary degree. We confine ourselves
to only two results. The first result consists of the two subleading
corrections to the first excited state of the cubic
potential,
\begin{equation}
\label{im3level1}
{\rm Im} \epsilon_1^{(3)}(g) = 
- \frac{ 8 {\rm e}^{-2/(15\,g)} }{ \sqrt{\pi} g^{3/2}} 
\left\{ 
1 - \frac{853}{16} g 
+ \frac{33349}{512} g^2 + \dots \right\},
\end{equation}
which have numerically large coefficients.
The second result concerns
the correction to the large-order growth of the 
perturbative coefficients for the ground state of the 
seventh-degree potential,
\begin{align}
\label{LO7}
& \epsilon_{0,K}^{(7)} = 
- \frac{ 5 \, \Gamma\left(5 K +\half \right)}{ 2^{2K+1} \, \pi^{3/2}} \, 
\left( \frac{18 \, \pi \, \sqrt{\phi} }%
  { 5^{1/4} \, \Gamma(\tfrac{1}{5}) \, 
    \Gamma^2(\tfrac{2}{5})} \right)^{5 K + 1/2} 
\nonumber\\[0ex]
& \quad \times \left\{ 1 - \frac{2^{1/2}\, 17\, \pi}%
  {5^{1/4}\, \phi^{3/2} \, 450\, K} + \dots \right\} \,.
\end{align}
Note that the absence of the $g^{1/2}$ correction from the 
result (\ref{A6}) implies the peculiar absence of a $1/K$ correction
to the leading factorial growth of perturbative coefficients for the 
sixth-degree potential (an observation which has been verified 
against high-precision numerical calculations~\cite{JeSuZJ2009prep}).

{\em Conclusions.---}
The general non-analytic expansions (\ref{evenodd}) for the 
resonance energies of even and odd oscillators
are triple expansions in terms of non-analytic exponentials,
logarithmic factors and fractional power series.
They follow from the 
generalized quantization conditions (\ref{EvenQuant}) and (\ref{OddQuant}).
The general leading-order behaviour of perturbation
theory for arbitrary levels of odd oscillators 
is given in Eq.~(\ref{leadingOdd}). These results allow
us to describe the widths of arbitrary 
resonances accurately by higher-order analytic formulas.
E.g.,~the first two corrections terms
given in (\ref{im3level1}) are indispensable for 
obtaining satisfactory agreement of the analytic
formula for the first excited cubic resonance energy 
with numerical calculations, even at small $g \approx 0.01$.

In a wider context, the following applications of our results can be envisaged:
in field theory (large-order estimates), the perturbations about the instanton
configurations have usually been neglected in the calculation of $n$-point
functions which enter the renormalization-group equations; our results indicate
that numerically large corrections may enter already on the level of model
calculations, and it may thus be worthwhile to revisit the subject.  Our
analysis may also be helpful for the analytic description of resonances in
quantum dot potentials which approximate the cubic anharmonic oscillator and
are important for quantum computing; from our analysis, it is obvious that in a
general potential, allowance should be made for higher-order correction terms
in the description of the width (a functional form 
$\exp(-A/|g|^p) \, [1 + B\, g +
C\, g^2 + \dots]$ with nonvanishing $B$ and $C$ seems to be indispensable).
From a more fundamental point of view, we can say that in potentials which
allow for tunneling of the particle, our analysis suggests that the familiar
Bohr--Sommerfeld--Wilson
quantization condition $\oint p\,\dd q = 2 \, \pi \, n\, \hbar$
still holds, but only if we assume a rather complicated analytic form for the
left-hand side of this quantization condition [see Eq.~\eqref{genie}].

U.D.J. acknowledges support from the Deutsche Forschungsgemeinschaft
(Je285/3-2) and helpful conversations with C. M. Bender, and A.S. 
acknowledges support from the Helmholtz Gemeinschaft (VH--NG--421).

\bibliographystyle{myprsty}

\end{document}